\documentclass[%
 reprint,
 amsmath,amssymb,
 aps,
]{revtex4-1}

\usepackage{graphicx}
\usepackage{dcolumn}
\usepackage{bm}

\newcommand{\calC}{{\mathcal{C}}}
\newcommand{\TBKTh}{T_{\mbox{\scriptsize BKT}}^{\mbox{\scriptsize high}}}
\newcommand{\TBKTl}{T_{\mbox{\scriptsize BKT}}^{\mbox{\scriptsize low}}}

\begin{document}


\title{Response to a twist on systems with $Z_{p}$ symmetry}

\author{Yuta Kumano$^{1}$}
\author{Koji Hukushima$^{2}$}
\author{Yusuke Tomita$^{3}$}
\author{Masaki Oshikawa$^{1}$}
\affiliation{$^{1}$Institute for Solid State Physics, University of Tokyo, Kashiwa 277-8581, Japan\\
$^{2}$Department of Basic Science, University of Tokyo, Tokyo 153-8902, Japan\\
$^{3}$College of Engineering, Shibaura Institute of Technology, Saitama 337-8570, Japan
}%
\date{\today}

\begin{abstract}
We study response to a twist in the two-dimensional $p$-state clock model,
which has the discrete $Z_{p}$ symmetry.
The response is measured in terms of helicity modulus, which
is usually defined with respect to an infinitesimal twist.
However, we demonstrate that such a definition is inappropriate
for the clock model.
The helicity modulus must be defined
with respect to a finite, quantized twist which matches the
discrete $Z_p$ symmetry of the model.
Recent numerical results, which casted doubt on the standard
picture that two Berezinskii-Kosterlitz-Thouless
transitions occur for $p > 4$, are rather a consequence of use
of the inappropriate quantity.
Numerical calculation of the appropriately defined helicity
modulus indeed supports the standard picture.
\end{abstract}

\pacs{Valid PACS appear here}
\maketitle


\textit{Introduction.---}
Symmetries govern various phases of matter and transitions
among them.
There are two distinct classes of symmetries: continuous and discrete,
with quite different consequences.
For example, gapless Nambu-Goldstone modes are required in
a spontaneous symmetry breaking phase, only if the symmetry is continuous.
The cyclic group $Z_p$ symmetry is discrete for any value of $p$.
Nevertheless, it approaches to the continuous U(1) symmetry in the
limit of $p\to \infty$.
It is thus an interesting problem to investigate how the phase
transition evolves as $p$ is increased.

Phase transitions in classical spin models, such as the classical XY
model, in two dimensions
with the continuous U(1) symmetry are an intriguing problem by itself.
Berezinskii-Kosterlitz-Thouless (BKT) transition~\cite{Berezinskii,KT}
is one of the most important concepts in statistical physics with a wide
range of implications.
The transition cannot be understood in terms of
spontaneous symmetry breaking.
Yet, the response to a twist, which is quantified
by the helicity modulus~\cite{Fisher},
clearly distinguishes the two phases separated
by the BKT transition: it is finite in the low-temperature
(BKT) phase while it is zero in the high-temperature (disordered) phase.
The helicity modulus exhibits a universal jump
at the BKT transition, which was confirmed numerically~\cite{Minnhagen}.
In the application to bosons in two dimensions,
the helicity modulus is identified with the superfluid density.
The universal jump of the helicity modulus implies the
universal jump in superfluid density~\cite{Nelson},
which was remarkably observed in experiments~\cite{Reppy}.

Let us now discuss the case with the discrete $Z_{p}$ symmetry.
As an explicit example, we consider the $p$-state clock model 
on the two-dimensional $L \times L$ square lattice with
periodic boundary conditions:
\begin{equation}
H_{p}= - \sum_{\langle \vec{r},\vec{r}' \rangle}\cos\left(\theta_{\vec{r}}
-\theta_{\vec{r}'} - A_{\vec{r},\vec{r}'} \right),
\label{eq.ham}
\end{equation}
where $\vec{r}=(r_1,r_2)$ 
($r_{1,2} \in \mathbb{Z}, 0 \leq r_{1,2} < L$),
$\langle \vec{r},\vec{r}' \rangle$ runs over all
the nearest neighbor sites on the square lattice,
and $\theta_{\vec{r}}$ takes integral multiples of $2\pi/p$.
We have set the coupling constant to unity.
The gauge field $A_{\vec{r},\vec{r}'}$
is usually set to zero but is introduced
to impose a twist, as we will discuss later.

According to the seminal renormalization group (RG) analysis~\cite{Jose}, 
for $p\leq 4$, there is a single critical point separating disordered and 
ordered phases, while the BKT phase appears for $p> 4$
in a finite range of temperature between the disordered
and ordered phases.
This has been also supported by later
examinations~\cite{Elitzur,Nomura,Ortiz}.
However, several recent numerical studies~\cite{Baek,Lapilli,Hwang,Baek_comment}
challenge the standard picture.
While details differ in each work, they suggest, in particular,
the absence of the BKT transition for $p=5$.
The most important evidence against the BKT transition picture is the
lack of the universal jump of the helicity modulus at the transition.

However, we will demonstrate that the apparent contradiction
between the numerical results and the RG picture
is rather an artifact due to the helicity modulus defined
inappropriately for the clock model.
Once defined properly, the behavior of the helicity modulus
is found to be perfectly consistent with the RG picture.

\textit{Definition of helicity modulus for systems with discrete symmetry.---}
The helicity modulus is often defined with respect to
an infinitesimal global twist.
We first introduce the gauge field
$A_{\vec{r}+\hat{x},\vec{r}} = \tilde{\Delta}/L$ for all the sites $\vec{r}$,
where $\hat{x}=(1,0)$ is the unit vector in the $x$ direction,
as a twist to the Hamiltonian~\eqref{eq.ham}. 
Then we define the helicity modulus by the second derivative 
$\tilde{\Upsilon}_{p} \equiv
\left. {\partial^{2} F_{p}(\tilde{\Delta})} /{\partial \tilde{\Delta}^{2}}
\right|_{\tilde{\Delta} = 0}$,
where $F_{p}(\tilde{\Delta})$ is the free energy in the presence of 
the twist $\tilde{\Delta}$ introduced above.
Being defined in terms of the derivative, it probes the response to
an infinitesimal twist spread over the entire system. 

This definition of the helicity modulus is convenient for numerical
calculations, since it can be written as an expectation value
of a local physical quantity:
\begin{equation}
 \tilde{\Upsilon}_{p} = \frac{1}{L^2} \left\langle \sum_{\vec{r}}
 \cos{ \phi_{\vec{r}+\hat{x},\vec{r}} } \right\rangle
- \frac{\beta}{L^{2}} \left\langle \left( \sum_{\vec{r}}
 \sin{ \phi_{\vec{r}+\hat{x},\vec{r}} }\right)^{2} \right\rangle
\label{eq.helicity.localform}
\end{equation}
where $\beta = 1/T$ is the inverse temperature
(we normalize the Boltzmann constant to unity), and
$\phi_{\vec{r}+\hat{x},\vec{r}}\equiv \theta_{\vec{r}+\hat{x}}-\theta_{\vec{r}}$.
We note that, for the infinitesimal global twist, the
resulting expression~\eqref{eq.helicity.localform} is
formally independent of $p$, and indeed the same
as in the XY model (which corresponds to $p \to \infty$).
However, the actual value of course depends on $p$,
since the expectation value is calculated for the $p$-state
clock model in which the variable $\theta$ is quantized to
integral multiples of $2\pi/p$.

It was indeed this definition of the helicity modulus that was used
in Refs.~\cite{Baek,Lapilli} to study the $p$-state
clock model.
The most striking aspect of their results for $p=5$ is that,
the helicity modulus thus obtained does not show the
universal jump which is expected for a BKT transition.
In fact, the helicity modulus is nonvanishing even in the
high-temperature disordered phase, in stark contrast to
the expected behavior at a BKT transition.
We also confirmed such behavior in our Monte Carlo simulations
for $p=5$.

This seems to contradict with the standard RG picture
with two BKT transitions~\cite{Jose}.
In the standard picture, the helicity modulus should
vanish in the disordered phase, while it diverges proportionally
to the system size $L$ in the ordered phase owing to the
extra free energy of the induced domain wall.
However, we find that $\tilde{\Upsilon}_{p}$ remaining nonzero
in the high-temperature phase is rather a consequence of
the inappropriate definition of the helicity modulus.
Indeed, for any finite $p$, we shall demonstrate that
the helicity modulus as defined in Eq.~\eqref{eq.helicity.localform}
is nonvanishing at arbitrary high temperature,
where the system is certainly disordered,
even in the thermodynamic limit.

To see this, let us consider the high-temperature (character) expansion
of Eq.~\eqref{eq.helicity.localform}, using the identity
$e^{\beta \cos\phi} = \sum_{k = -\infty}^{\infty} I_k(\beta) e^{i k \phi}$
for the modified Bessel function $I_k$.
For the XY model, we find
\begin{equation}
\tilde{\Upsilon} = \frac{1}{\beta L^2}
\sum_{\substack{ \{ k_{ \vec{r}, \vec{r}'} \in \mathbb{Z}  \} \\
\sum_{\vec{r}'} k_{\vec{r},\vec{r}'} = 0 }}
\left(\sum_{\vec{r}} k_{\vec{r},\vec{r}+\hat{x}}\right)^2
\prod_{ \langle \vec{r},\vec{r}' \rangle }
\frac{I_{k_{\vec{r},\vec{r}'}}(\beta)}{I_0(\beta)},
\label{eq.helicity.highTexp}
\end{equation}
where $k_{\vec{r},\vec{r}'}$ is an integer defined for each link between
the nearest neighbor pairs, and
$k_{\vec{r}',\vec{r}} \equiv - k_{\vec{r},\vec{r}'}$.
The constraint $\sum_{\vec{r}'} k_{\vec{r},\vec{r}'} = 0$ means that
the lattice divergence vanishes at every site.
Thus, each term in the high-temperature expansion corresponds
to a configuration of an integer-valued
``flux field'' $\{ k_{\vec{r},\vec{r}'} \}$ forming closed loops.
Since $I_k(\beta) \sim \beta^{|k|}$,
each term corresponds to the order
$\sum_{\langle \vec{r}, \vec{r}' \rangle} | k_{\vec{r},\vec{r}'} |$
of the high-temperature expansion.
Unless there is a flux loop winding the entire system,
which can occur only above the order $L$,
we have
$\sum_{\vec{r} } k_{\vec{r},\vec{r}+\hat{x}} = 0$.
Thus, for the XY model in the thermodynamic limit $L \to \infty$,
the helicity modulus defined in
Eq.~\eqref{eq.helicity.localform} vanishes
at any finite order in the high-temperature
expansion~\eqref{eq.helicity.highTexp}.

Now let us consider the $p$-state clock model. The only
difference from the XY model comes from the fact that
$\sum_{\theta = 0, 2\pi/p, \ldots, 2\pi(p-1)/p} e^{i n \theta} = p$,
if $n$ is an integral multiple of $p$.
Thus, the high-temperature expansion of Eq.~\eqref{eq.helicity.localform}
for the $p$-state clock model is exactly the same
as Eq.~\eqref{eq.helicity.highTexp}, except for the constraint
at each site now replaced by
$\sum_{\vec{r}'} k_{\vec{r},\vec{r}'} \equiv 0 \mod{p}$.
This allows various flux configurations which were not permitted
in the XY model.
The lowest order in the high-temperature expansion among such
configurations is given by $k_{\vec{R},\vec{R}+\hat{x}} = p$
and all other fluxes being zero $k_{\vec{r},\vec{r}'}=0$.
This leads to the nonvanishing contribution
\begin{equation}
\tilde{\Upsilon}_{p} \sim \frac{p}{(p-1)!2^{p-1}}\beta^{p-1}
\end{equation}
in the lowest order of $\beta$.
This proves that the helicity modulus as defined
in Eq.~\eqref{eq.helicity.localform}
does not vanish at arbitrary high temperature, for any finite $p$, 
even in the thermodynamic limit.
This implies that Eq.~\eqref{eq.helicity.localform} does
not sharply distinguish the disordered and other phases.
We note that, while this holds for an arbitrary large finite $p$,
the contribution becomes rapidly smaller for larger $p$.
This is the reason why the problem in the high-temperature phase
has been noticed numerically only for small $p$, especially
$p=5$.

Thus we need a different quantity to describe the phase transition.
Intuitively, the problem with the infinitesimal twist we have used
could be understood as the introduction of a mismatch between neighboring
spins, whose directions are quantized in the $p$-state clock model.
Therefore, let us define the helicity modulus as a response to
the finite twist
$A_{\vec{r}+\hat{x},\vec{r}} = \Delta$
localized on the horizontal links localized on a single
column $\vec{r} \in \calC \equiv \{ (L-1,r_2) |  0 \leq r_2 < L \}$:
\begin{equation}
\Upsilon_{p} = \frac{2(F_{p}(\Delta) - F_{p}(0))}{\Delta^{2}}.
\label{eq.discrete.helicity}
\end{equation}
Matching with the $Z_p$ symmetry of the model would require
$\Delta$ to be an integral multiple of $2\pi/p$.
We note that, in the original introduction of
the helicity modulus~\cite{Fisher}, the twist was not restricted to
be infinitesimal.
Furthermore, once inside the BKT phase, the helicity modulus
thus defined should behave in the same way as
the one with respect to an infinitesimal twist.

In the high-temperature expansion,
the difference between $Z_{p}(\Delta)$ and $Z_{p}(0)$
comes only from the factor
$\exp{\big[ i \Delta \sum_{\vec{r} \in \calC}
k_{\vec{r},\vec{r}+\hat{x}} \big]}$.
Thus, if $\Delta$ is an integral multiple of $2\pi /p$,
the effect of the twist in the $p$-state clock model
again disappears, even under the relaxed constraint
$\sum_{\vec{r}'} k_{\vec{r},\vec{r}'} \equiv 0 \mod{p}$,
at any finite order of the high-temperature expansion
in the thermodynamic limit.
The helicity modulus thus defined indeed
satisfies the physical expectation,
vanishing exactly in the high-temperature disordered phase.


\textit{Numerical calculations of the free energy difference due to a twist.---}
\begin{figure*}
\includegraphics{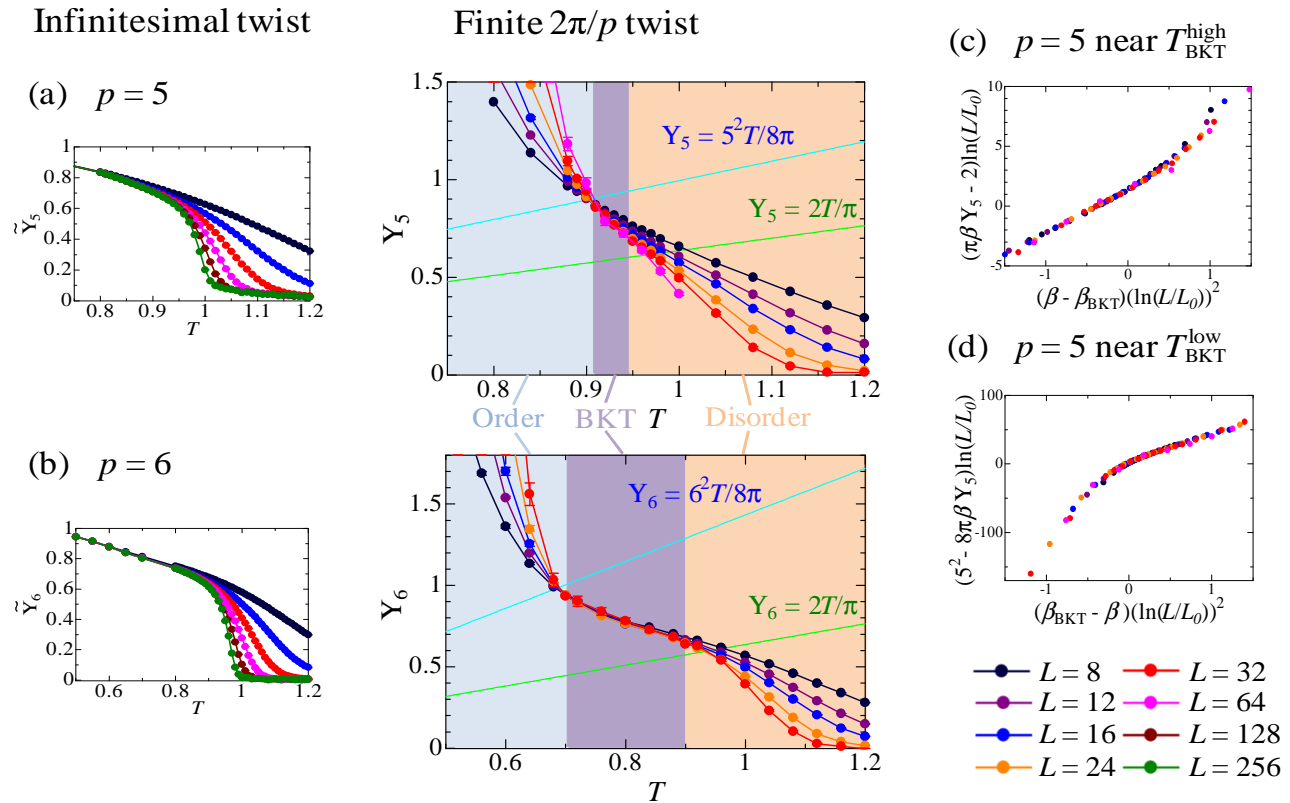}
\caption{(Color online) (a),(b) Temperature dependence of $\tilde{\Upsilon}_{p}$
and $\Upsilon_{p}$ for $p=5$ and $p=6$.
In both cases, $\tilde{\Upsilon}_{p}$ with respect to
a infinitesimal twist does not reflect each phase appropriately.
In comparison, for $\Upsilon_{p}$ with respect to the finite twist $2\pi/p$,
 we can find three regions, blue, purple and beige regions, 
correspond to ordered, BKT and disordered phase, respectively.
In addition, as $L$ is increased, $\Upsilon_{p}$ approaches
 the universal value on the higher and lower transition points, 
with the universal relations~\eqref{eq.jump.high} and~\eqref{eq.jump.low}
shown respectively by yellow-green and aqua lines.
Colored lines are for a guide to the eyes.
(c),(d) Finite-size scaling of $\Upsilon_{5}$ near $\TBKTh$ and $\TBKTl$.
All the data points collapse into universal curves in both cases. 
The estimated scaling parameters are $(T_{\rm{BKT}}, L_{0}) = (0.944, 0.8),(0.908, 1.9)$
 for (c) and (d), respectively.}
\label{fig.p5p6andFSS}
\end{figure*}
We have concluded that, in order to distinguish different phases of 
the clock model in terms of its response to a twist, the twist angle must be 
an integral multiple of $2\pi/p$.
The change in the free energy due to a twist of a finite angle cannot be 
simply reduced to an expectation value of a local physical quantity, 
unlike in the case of the infinitesimal twist.
Thus it is more difficult to calculate the response in Monte Carlo simulations. 
Nevertheless, it is possible to do so, as we demonstrate in the following.
What we have to calculate is the ratio of the partition functions,
$Z_{p}(\Delta)/Z_{p}(0)$, 
and this quantity can be calculated by boundary-flip Monte
carlo method~\cite{Hasenbusch,Hukushima}.
To overcome the critical slowing down of spin variables,
we use Wolff cluster method~\cite{Wolff} at the low temperature.
In our simulations, the number of Monte Carlo step are $O(10^{7})$ up to 
$O(10^{9})$ after equilibration.

We first studied the helicity modulus~\eqref{eq.discrete.helicity}
defined with respect to the finite quantized twist $2\pi/p$
for $p \leq 4$, where a single transition due to the spontaneous
symmetry breaking is expected.
We indeed find, as the system size is increased,
$\Upsilon_{p} \to 0$ in the high-temperature disordered
phase and $\Upsilon_{p} \propto L$ in the low-temperature ordered phase.
The known critical temperatures for $p=2,3,$ and $4$
are reproduced from the change in the behavior of $\Upsilon_{p}$
(data not shown in this Letter).

Now let us come to the main question of $p> 4$, where
the standard RG picture predicts two BKT transitions
while some of the recent numerical studies found
apparent contradictions with it.
We emphasize that, while those recent papers do
agree with the RG picture
and identify BKT transitions for sufficiently large $p$,
the helicity modulus
as defined in Eq.~\eqref{eq.helicity.localform} they studied
is non-zero at arbitrary high temperature.
It is just that Eq.~\eqref{eq.helicity.highTexp} is smaller
for higher values of $p$
and more difficult to be detected numerically.
We still need to verify if the alternative
definition, Eq.~\eqref{eq.discrete.helicity}, indeed
confirms the BKT transitions.
We show the temperature dependence of
the helicity modulus $\Upsilon_{p}$, Eq.~\eqref{eq.discrete.helicity},
with respect to the finite twist, for $p=5$ and $6$,
in Fig.~\ref{fig.p5p6andFSS}(a) and~\ref{fig.p5p6andFSS}(b).
In the low-temperature regime, colored blue,
$\Upsilon_{p}$ increases proportionally to the system size $L$,
while it decreases towards zero in the high-temperature regime,
colored beige.
These are precisely the expected behavior in the low-temperature
ordered (spontaneously symmetry breaking)
and the high-temperature disordered phases.
In contrast, the helicity modulus $\tilde{\Upsilon}_{p}$, defined
as Eq.~\eqref{eq.helicity.localform}  
with respect to the infinitesimal twist,
does not exhibit physically expected behaviors for a helicity modulus.
While the disagreement of Eq.~\eqref{eq.helicity.localform}
with the standard RG picture in
the high-temperature phase, especially for $p=5$,
was discussed previously,
we also emphasize that Eq.~\eqref{eq.helicity.localform}
does not detect the low-temperature ordered phase.
The ``finite twist'' helicity modulus~\eqref{eq.discrete.helicity}
we have introduced
does distinguish the three phases, as is physically expected. 

Moreover, in the intermediate temperature regime (purple),
$\Upsilon_{p}$ tends to be independent of $L$,
which is indeed the expected behavior in the BKT phase.
This is rather clear for $p=6$; the approximate transition
temperatures estimated from Fig.~\ref{fig.p5p6andFSS}(b) for $p=6$, 
$\TBKTh \sim 0.90$ and $\TBKTl \sim 0.70$, are in
agreement with the estimates
$\TBKTh \sim 0.90008(6)$
and 
$\TBKTl \sim 0.7014(11)$
obtained by a different method~\cite{Tomita}.
On the other hand, the existence of the BKT phase
for $p=5$ is not very clear in Fig.~\ref{fig.p5p6andFSS}(a),
although it does not contradict the data.
In addition, it is desirable to obtain more quantitative
verification also for $p=6$.
Therefore we further examine our numerical data with
the BKT transition scenario.

The well-known consequence of the BKT transition picture
is the ``universal jump'' of the helicity modulus.
In terms of RG, it is a consequence of the fact that
the BKT transition occurs at a definite Gaussian coupling constant
(also known as ``Luttinger parameter'' or ``compactification radius'')
where the leading perturbation becomes marginal.
A similar argument can be also applied to the transition at the
low-temperature side of the BKT phase~\cite{Jose}.
For $p> 4$, we obtain the universal relations at the two transitions as
\begin{eqnarray}
\lim_{T\to T_{\mbox{\tiny BKT}}^{\mbox{\tiny high} }-0}
\Upsilon_{p} &=&
\frac{2}{\pi} \TBKTh
\label{eq.jump.high}
\\
\lim _{T\to T_{\mbox{\tiny BKT}}^{\mbox{\tiny low}}+0}
\Upsilon_{p} &=&  \frac{p^2}{8\pi} \TBKTl.
\label{eq.jump.low}
\end{eqnarray}
In Fig.~\ref{fig.p5p6andFSS}(a) and~\ref{fig.p5p6andFSS}(b),
we compare the temperature dependence of
the ``finite twist'' helicity modulus~\eqref{eq.discrete.helicity},
 with the universal relations~\eqref{eq.jump.high} and~\eqref{eq.jump.low}
shown respectively by  yellow-green and aqua lines.
We can find, as $L$ is increased, $\Upsilon_{p}$ approaches
 the universal values on the higher and lower transition points.
They do not, however, completely converge owing to finite-size effects.

Therefore, we have also performed a finite-size scaling analysis
based on the RG theory at each transition.
In the vicinity of the transition at higher temperature,
$\TBKTh$, the $Z_p$ anisotropy $y_{p}$ is always 
irrelevant, so we can approximately neglect $y_p$ to obtain
the standard BKT RG equations for the vortex fugacity $y$ and
the Gaussian coupling~\cite{Jose,Elitzur}.
The scaling form of the helicity modulus
was derived~\cite{Harada_Kawashima} from the RG equations as
\begin{equation}
x(T,L) = l^{-1} f( l^{2} \delta),
\label{eq.scaling}
\end{equation}
where $x=\pi \beta \Upsilon_{p} -2$,
$\delta = \beta -\beta_{{\rm BKT}}^{{\rm high}}$ and $l=\ln(L/L_{0})$.
Here, $\beta_{{\rm BKT}}^{{\rm high}} = 1/ \TBKTh$ is the inverse transition temperature,
and $L_{0}$ is a scaling parameter.
They are the fitting parameters to the data.
On the other hand, near the transition at the lower temperature,
$\TBKTl$, the vortex fugacity $y$ is always irrelevant
and thus may be neglected.
The RG equations involving $y_p$,
in the vicinity of the lower-temperature transition,
$\TBKTl$, are similar to that for the
higher-temperature transition at $\TBKTh$.
As a consequence, for the lower-temperature transition at $\TBKTl$,
the scaling equation is obtained from Eq.~\eqref{eq.scaling}
by replacing
$x$ and $\delta$ by $x'=p^{2} - 8\pi \beta \Upsilon_{p}$
and $\delta'= \beta_{{\rm BKT}}^{{\rm low}} - \beta$ respectively, where
$\beta_{{\rm BKT}}^{{\rm low}} = 1 / \TBKTl$,
and the scaling function $f$ by a different scaling function $f'$.

In Fig.~\ref{fig.p5p6andFSS}(c) and~\ref{fig.p5p6andFSS}(d),
we show the finite-size scaling of $\Upsilon_{5}$ near 
each of the transitions, at $\TBKTh$ and $\TBKTl$.
We can find all the data points collapse into universal functions 
if we set two scaling parameters as
$(T_{\rm{BKT}}, L_{0}) = (0.944, 0.8), (0.908, 1.9)$, respectively.
These are in good agreement with $\TBKTh = 0.95147(9)$
 and $\TBKTl = 0.90514(9)$ 
estimated earlier by a different numerical method~\cite{Papa}.

We have also confirmed, for $p=6$,
that $\Upsilon_{6}$ has the universal value
at each BKT transition point.
We estimate the transition temperatures $\TBKTh = 0.904(2)$ 
and $\TBKTl = 0.700(2)$ by the finite-size scaling~\eqref{eq.scaling}.
These are in a very good agreement with $\TBKTh = 0.90008(6)$
and $\TBKTl = 0.7014(11)$ determined earlier by a different
numerical method~\cite{Tomita}.

\textit{Conclusions and discussions.---}
All our observations support, for $p> 4$,
the existence of the BKT phase separated by two BKT transitions from
the high-temperature disordered phase and from the
low-temperature ordered phase, as predicted by RG~\cite{Jose}.
Much of the apparent contradiction with recent numerical results
was rather due to the ``helicity modulus''
adopted in the numerical studies, which was defined with respect
to an infinitesimal twist.
Such a quantity is inappropriate for the $p$-state clock model
with the discrete $Z_p$ symmetry.
We have established that, for this model,
the helicity modulus instead has to be defined with respect to
a finite quantized twist which matches the $Z_p$ symmetry
of the system.

Although we have restricted the analysis
to the simple $p$-state clock model in the present
Letter, our discussion can be readily applied to
more general models with $Z_p$ symmetry, including
the XY model with a perturbation which breaks the
U(1) symmetry down to $Z_p$.
The present result would also have implications in a wider
range of problems with discrete symmetries or
discretized degrees of freedom.


\medskip

We would like to thank Naoki Kawashima for valuable suggestions 
and fruitful discussions.
The present work is supported in part by
Grant-in-Aid for Scientific Research (KAKENHI)
for Priority Areas No. 20102008 from MEXT of Japan,
and by US National Science Foundation
Grant No. NSF PHY11-25915 while Y.~K. and M.~O. performed
a part of the present work at Kavli Institute for
Theoretical Physics, UC Santa Barbara.
Y.~K. was partially supported by Advanced Leading Graduate Course 
for Photon Science (ALPS) grant.
A part of numerical calculations for this work was done at
Supercomputer Center, Institute for Solid State Physics,
University of Tokyo.

\nocite{*}

\bibliography{ref}

\end{document}